\documentclass[aps,prb,twocolumn,a4paper,groupedaddress]{revtex4-1}
\usepackage{amsmath}
\usepackage{amssymb}
\usepackage{bm}
\usepackage{color}
\usepackage[dvips]{graphicx}

\definecolor{myorange}{rgb}{0.7,0.5,0.0}
\definecolor{mygreen}{rgb}{0.0,0.7,0.0}
\definecolor{mymagenta}{rgb}{0.7,0.0,0.7}

\def\ket#1{\mathinner{|{#1}\rangle}}
\def\braket#1{\mathinner{\langle{#1}\rangle}}

\begin{document}
  \title{Competition among  various charge-inhomogeneous states and {$d$-wave superconducting state} in Hubbard models on square lattices}
  \author{Kota Ido, Takahiro Ohgoe and Masatoshi Imada}
  \affiliation{Department of Applied Physics, University of Tokyo, 7-3-1 Hongo, Bunkyo-ku, Tokyo 113-8656, Japan}

\begin{abstract}
We study competitions between charge uniform and inhomogeneous states in two-dimensional Hubbard models 
by using a variational Monte Carlo method. 
At realistic parameters for cuprate superconductors, emergent effective attraction of carriers generated from repulsive Coulomb interaction leads to
charge/spin stripe ground states, which severely compete with uniform  superconducting excited states in the energy scale of 10 K for the cuprates.
Stripe period increases with decreasing hole doping $\delta$, which agrees with the experiments for La-based cuprates at $\delta=1/8$. 
For lower $\delta$, we find a phase separation.
Implications of the emergent attraction for the cuprates are discussed.
\end{abstract}

\maketitle
%
\section{Introduction}
%
After the discovery of the high temperature superconductivity in the cuprates\cite{Bednorz1986}, its mechanism remains one of the most challenging issues in condensed matter physics.
A necessary condition of high-temperature superconductivity for strongly correlated electron systems is a large effective attractive interactions between electronic carriers emerging from strong Coulomb repulsions.
However, this strong attraction can also enhance the tendency of electron aggregations in real space.
This means that the strong attractive interaction induces diverging charge compressibility~\cite{Furukawa1992,Furukawa1993} as well as charge inhomogeneous states such as phase separations (PS) and stripe states\cite{Zaanen1989,Emery1990,Kivelson1996,White2000,Himeda2002,Corboz2011,Scalapino2012,Corboz2014,Capone2006,Misawa2014a,Otsuki2014,Zhao2017,Ido2017}.
In fact, the competition between the superconductivity and the charge inhomogeneity as a stripe state has been observed and well discussed in La-based cuprates\cite{Tranquada1995,Tranquada1996,Yamada1998,Hucker2013,Fink2011}.
Recently, such phenomena were also reported in Y-\cite{Ghiringhelli2012,Tabis2014,Comin2015,Comin2015a,Forgan2015,Peng2016}, Hg-\cite{Tabis2014,Campi2015} and Bi-based cuprates\cite{Mesaros2016a,Eduardo2014,Fujita2014}, 
indicating a ubiquitous feature in the cuprate superconductors\cite{Keimer2015,Comin2016}.

To understand the origin of superconductivity in the cuprates, the Hubbard model on a square lattice has been studied for long time.
Although many theoretical studies have been devoted to understanding the ground states of the Hubbard model, they are still under debate\cite{Giamarchi1991,Furukawa1992,Yokoyama2013,Capone2006,Aichhorn2007,Kancharla2008,Chang2008,Chang2010,Khatami2010,Sordi2012,Gull2012,Misawa2014a,Otsuki2014,Zheng2015,Leblanc2015,Tocchio2013,Zheng2017,Zhao2017}.
To gain insight into the charge inhomogeneous phases including the stripes, 
detailed analyses of their existence and competitions with the $d$-wave superconductivity are desired, particularly on their dependences on the hole doping concentration $\delta$, band structure and the interaction. 
Most numerical studies based on variational calculations or dynamical mean-field theory showed that charge uniform states are the ground states or macroscopic phase separation appears\cite{Yokoyama2013,Capone2006,Aichhorn2007,Kancharla2008,Khatami2010,Sordi2012,Misawa2014a,Tocchio2016}. 
However, in these calculations, the possibility of long-period stripe states are ignored.
Recent studies using infinite projected entangled pair states, the density matrix embedding theory (DMET), constrained path auxiliary field quantum Monte Carlo method and density matrix renormalization group all reported the stripe ground state, but studied systematically only for a special choice of band structure (only with nearest neighbor transfer $t=1$) at $\delta=0.125$, with $8$/$16$ period for charge/spin stripes\cite{Zheng2017}.
Recent variational Monte Carlo (VMC) calculations combined with tensor network states also found stripe states with  $8$/$16$ (for $\delta<0.15$) and $4$/$8$ (for $\delta>0.15$) periods for charge/spin as ground states below $\delta \sim 0.25$\cite{Zhao2017}.
However, the stripe period extensively studied at $\delta=0.125$ in these calculations is different from that observed in La-based cuprates, which is $4$ charge and $8$ spin periods\cite{Tranquada1995,Tranquada1996}.
These results imply that more systematic and realistic study is needed to understand the real cuprate systems.

One of the missing ingredients in the simple Hubbard model is hopping parameters beyond the nearest-neighbor pairs.
The previous DMET study showed that the stripe state in the experiments has a lower energy than the charge uniform state in the system with the next-nearest hopping\cite{Zheng2015}.
However, since the sizes of embedded clusters are restricted, the competitions with other stripe states are still unclear at a finite hole concentration. 

In this paper, by using the VMC method, 
we study the competitions among stripe states with different periodicities in addition to charge uniform states. 
We show that the ground states has stripe orders, the period of which decreases with increasing $\delta$ in a wide range.
In the lower doping region, the PS occurs between the antiferromagnetic insulator and the stripe state.
More importantly, we find that the stripe state experimentally observed at $\delta=0.125$ is indeed the ground state for a realistic value of next-nearest-neighbor hopping.
We clearly see that the superconducting (SC) long-range order is strongly suppressed due to the emergence of stripe orders, while charge uniform and strong superconducting states exist as excited states with tiny excitation energies.

%
\section{Model and Method}
%
We study $t-t'$ Hubbard model on square lattices under the antiperiodic-periodic boundary condition. 
The Hamiltonian is defined by 
\begin{eqnarray}
  \mathcal{H} =&& -\sum_{i,j,\sigma} t_{ij}c^\dagger_{i\sigma} c_{j\sigma}+ U\sum_i^{N_s} n_{i\uparrow}n_{i\downarrow},
\end{eqnarray}
where the hopping amplitude $t_{ij}$ is taken as $t_{ij}=t$ for the nearest-neighbor pairs, $t_{ij}=t'$ for the next-nearest-neighbor pairs and otherwise $t_{ij}=0$.
$U$ is the onsite repulsive interaction, $N_s = L \times L$ is the system size, $c^\dagger_{i\sigma}$ ($c_{i\sigma}$) is a creation (annihilation) operator of an electron with spin $\sigma$ on the site $i$, and $n_{i\sigma} = c^\dagger_{i\sigma}c_{i\sigma}$. 
The lattice constant is taken as the length unit.
We mainly performed the calculations for $U/t=10$ because it is close to proposed {\it ab initio} estimate for the cuprates~\cite{HirayamaMisawaYamajiImada2017}.

To study the ground states of the Hubbard model, we have used the VMC method.
As a trial wave function, we adopted the generalized pair product wave function with correlation factors: $\ket{\psi} = \mathcal{P}_G \mathcal{P}_J \mathcal{P}_{\rm d-h}^{\rm ex} \ket{\phi}$~\cite{Tahara2008a}. 
Here Gutzwiller factor  $\mathcal{P}_G=\exp \left( -g\sum_i n_{i\uparrow}n_{i\downarrow}\right)$, 
Jastrow factor $\mathcal{P}_J=\exp \left( -\sum_{i,j}v_{ij} n_{i}n_{j}\right)$, and
the doublon-holon correlation factor  $\mathcal{P}_{\rm d-h}^{\rm ex}=\exp \left( -\sum_{m=0}^5 \sum_{l=1,2} \alpha_{(m)}^{(l)} \sum_i \xi_{i(m)}^{(l)}\right)$ are considered and 
 $\ket{\phi} = \left( \sum_{i,j}^{N_s} f_{ij} c_{i\uparrow}^\dagger c_{j\downarrow}^\dagger \right)^{N/2} \ket{0}$, where $n_i=n_{i\uparrow}+n_{i\downarrow}$ and $N$ is the number of electrons. 
 $\xi_{i(m)}^{(l)}$ is 1 when a doublon (holon) exists at the $i$-th site and $m$ holons (doublons) surround at the $l$-th nearest neighbor.
Otherwise, $\xi_{i(m)}^{(l)}$ is 0.
In this study, we treat $g,v_{ij},\alpha_{(m)}^{(l)}$ and $f_{ij}$ as variational parameters. 
To describe inhomogeneous stripe states, we assume that $f_{ij}$ has the $l_s \times 2$ sublattice structure, which enables the $l_s$ period spin stripe.
In our calculations, we treat several tens of thousands of variational parameters for the largest systems.
All the variational parameters are optimized by using the stochastic reconfiguration method\cite{Sorella2001}.

To clarify physical properties of the ground states, we measured the spin structure factor $S_{\rm s}(\bm{q})=\frac{1}{3N_s} \sum_{i,j} \braket{\bm{S}_i \cdot \bm{S}_j}e^{-i\bm{q}\cdot(\bm{r}_i-\bm{r}_j)}$, 
the charge structure factor $S_{\rm c}(\bm{q})=\frac{1}{N_s} \sum_{i,j} \braket{n_i n_j-\rho^2}e^{-i\bm{q}\cdot(\bm{r}_i-\bm{r}_j)}$ and the long-range part of $d_{x^2-y^2}$-wave SC correlation functions $P_d^{\infty}=\frac{1}{M}\sum_{|\bm{r}| \geq r_{\rm max}/2} P_d(\bm{r})$, 
where $M$ is the number of vectors satisfying $|\bm{r}| \geq r_{\rm max}/2$.
Here, $\rho=\sum_{i,\sigma}\braket{n_{i\sigma}}/N_s$, $r_{\rm max}=L/\sqrt{2}$ and $P_d(\bm{r})=\frac{1}{2N_s}\sum_i \braket{\Delta^\dagger_d(\bm{r})\Delta_d(\bm{r}+\bm{r}_i) + \Delta_d(\bm{r})\Delta^\dagger_d(\bm{r}+\bm{r}_i)}$ with $\Delta_d(\bm{r}_i)=\frac{1}{\sqrt{2}}\sum_{\bm{r}} g(\bm{r})(c_{\bm{r}_i\uparrow}c_{\bm{r}_i+\bm{r}\downarrow} - c_{\bm{r}_i\downarrow}c_{\bm{r}_i+\bm{r}\uparrow}) $. 
The form factor $g(\bm{r})$ is defined as $g(\bm{r})=\delta_{r_x,0}(\delta_{r_y,1}+\delta_{r_y,-1})-\delta_{r_y,0}(\delta_{r_x,1}+\delta_{r_x,-1})$, where $\bm{r}=(r_x,r_y)$. 
We define the spin/charge order parameter as $\Delta_{\rm S/C}=\sqrt{S_{\rm s/c}(\bm{q}_{\rm peak})/N_s}$, where $S_{\rm s/c}(\bm{q}_{\rm peak})$ represents the peak value of the spin/charge structure factor. 
We also define the SC order parameter as $\Delta_{\rm SC}=\sqrt{P_d^{\infty}}$.

%
\section{Results}
%
%
\subsection{Ground-state phase diagram of the $t-t'$ Hubbard model}
\begin{figure}[htbp]
  \begin{center}
   \includegraphics[width=82mm]{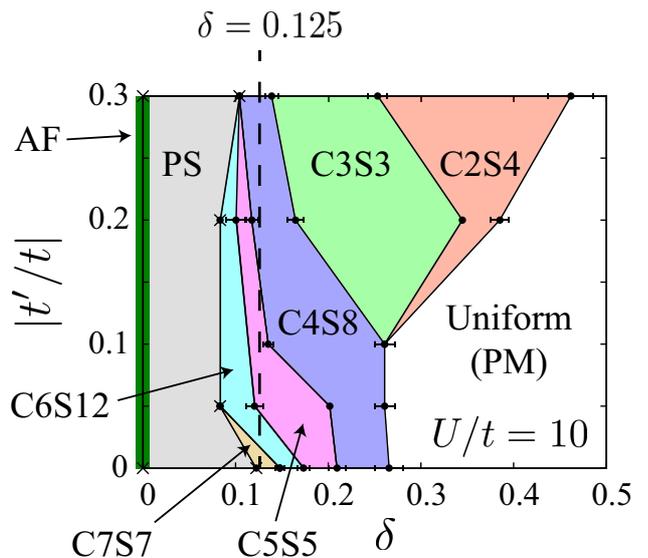}
  \end{center}
  \caption{
    (Color online) Ground-state phase diagram of the Hubbard model on a square lattice for $U/t=10$.
    Note that $t'/t$ is a negative value.
    At $\delta=0$, the ground state is the antiferromagnetic (AF) Mott insulator (green bold line). 
    Cross symbols indicate the boundary of the phase separation (PS).
    Solid black circles represent the boundaries of ``C$l_c$S$l_s$" stripe states with $l_c$/$l_s$ period for charge/spin.
    Dashed line shows $\delta=0.125$.
    Solid lines and painted regions are guides for the eyes.
    In the unpainted (white) region, the ground state is a charge uniform paramagnetic (PM) state.
  }
  \label{pd}
\end{figure}

%
The main results are summarized in Fig. \ref{pd}, which shows the ground-state phase diagram in the $\delta-t'$ plane for $U/t=10$.
Throughout this paper, the stripe state with charge (spin) period $l_c$($l_s$) is denoted as ``C$l_c$S$l_s$'' for simplicity. 
Charge uniform states are obtained under the 2$\times$2 sublattice structures and energies are compared with inhomogeneous states obtained under longer sublattices.
As shown in Fig. \ref{pd},  charge inhomogeneous states exist as the ground states in a wide range of $\delta$ for any $t'/t$.
The wavelength of the charge $l_c$ becomes longer with the decrease of $\delta$, and eventually the PS, whose wavelength is infinite, occurs between the antiferromagnetic insulator and a stripe state. 
For $-0.3 \leq t'/t \lesssim -0.15$, which is a realistic range of $t'/t$ for the cuprates, the ground state at $\delta=1/8$ is the C4S8 state which has been observed in {\rm La}-based cuprates\cite{Tranquada1995,Tranquada1996}.
However, charge inhomogeneous states are stabilized even in the highly overdoped regime and thus a uniform $d$-wave superconducting state does not appear as the ground state of the single-band Hubbard model {\it at strong coupling}. 
We will discuss our numerical results in comparison with the experiments in Sec. \ref{discussion}.

%
\subsection{Ground states and excited states}
%
First, we show results for $t'/t = 0$ as a simplest model.
\begin{figure}[htbp]
  \begin{center}
   \includegraphics[width=82mm]{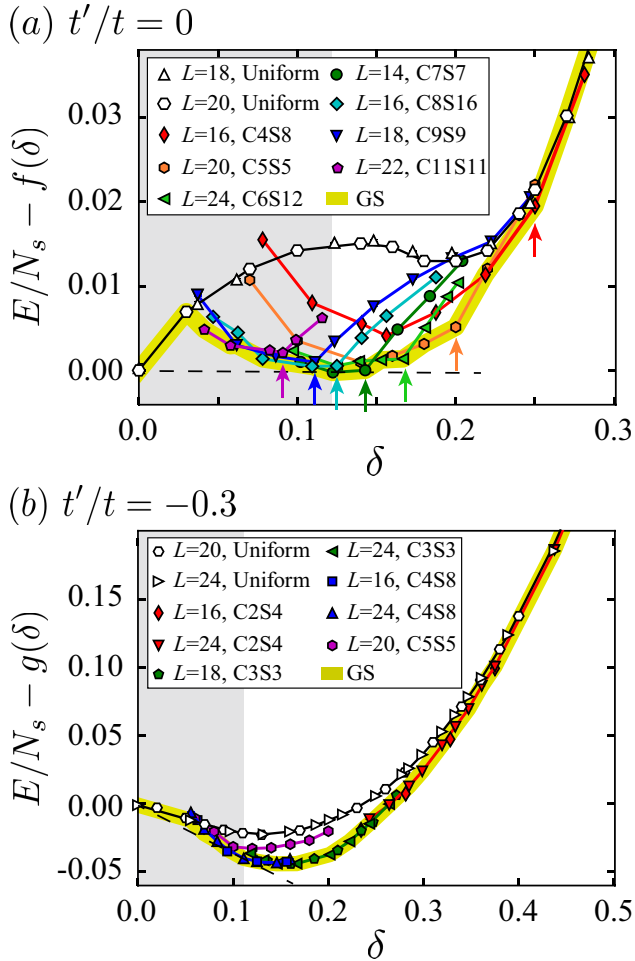}
  \end{center}
  \caption{(Color online) Doping concentration dependence of energies for several different states in two-dimensional Hubbard model with $U/t=10$ at (a) $t'/t=0$ and (b) $t'/t=-0.3$.
    A linear function $f(\delta)=-1.835\delta-0.4211$ or $g(\delta)=-1.5\delta-0.4222$ is subtracted for better visibility.
  For clarity, we draw yellow thick line to represent the energies of the ground states. 
  Types of states and system sizes are described in the legend.
  Error bars indicate the statistical errors arising from the Monte Carlo sampling, but most of them are smaller than the symbol sizes here and in the following figures.
  Dashed black line and gray region show the tangent line of the energy curve drawn from $\delta=0$ and PS, respectively.
  In panel (a), commensurate fillings $\delta=1/l_c$ are indicated by colored arrows.
  }
  \label{ene_tp0.0}
\end{figure} 
Figure \ref{ene_tp0.0} (a) shows the energies of uniform and stripe states with different periodicities as functions of hole-doping concentration $\delta = 1-N/N_s$. 
We will show evidences for the stripe long-range order described in Fig.~\ref{ene_tp0.0} (a) later in Fig.~\ref{phys_tp0.0}.
From Fig. \ref{ene_tp0.0} (a), we see that stripe states are the ground states below $\delta \approx 0.25$.
The maximum value of energy difference between uniform and stripe states is the order of $\sim0.01t$ at $\delta \approx 0.125$, which is consistent with the recent results by other numerical calculations such as the tensor network states\cite{Zheng2017,Zhao2017}.
By increasing the hole concentration, the wavelength of the charge $l_c$ becomes shorter. 
This is naturally related to the  mean distance between holes, which decreases with increasing doping concentrations.
Stripe states with $l_c \leq 3$ were not found as the ground states.

\begin{figure}[htbp]
  \begin{center}
   \includegraphics[width=82mm]{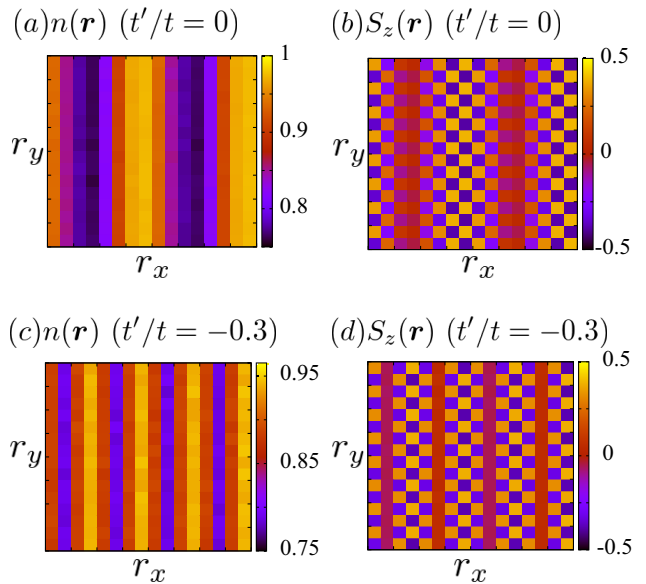}
  \end{center}
  \caption{
    (Color online) Charge density $n(\bm{r})=\braket{n_{\bm{r}\uparrow}+n_{\bm{r}\downarrow}}$ 
    and spin density along $z$-direction $S_z(\bm{r}) = 0.5\braket{n_{\bm{r}\uparrow}-n_{\bm{r}\downarrow}}$ 
    for the ground state for $L = 16$ and $U/t=10$ at $\delta = 0.125$.
    The next-nearest-neighbor hopping in (a-b) and (c-d) are $t'/t=0$ and $t'/t=-0.3$, respectively.
  }
  \label{config}
\end{figure}

To clarify the possibility of PS, we performed a Maxwell construction for the energy curve of the ground states (dashed line in Fig. \ref{ene_tp0.0} (a)).
We find that a PS appears for $0< \delta \leq 0.125$. 
This region is narrower than that obtained in the previous VMC study, where only uniform states were assumed\cite{Misawa2014a}.
Then we conclude that the stripe states are stable ground states in the region $0.125 < \delta < 0.25$. 
At $\delta \approx 0.125$, several stripe states for $l_c=6-8$ are nearly degenerate, which is also consistent with recent studies by state-of-the-art numerical methods\cite{Zheng2017}.
The charge and spin configurations of the C8S16 state at $\delta=0.125$ are plotted in Figs. \ref{config} (a) and (b), respectively.

Next, we show the results for $t'/t = -0.3$, which is a realistic value for the cuprate superconductors~\cite{HirayamaMisawaYamajiImada2017}.
Figure \ref{ene_tp0.0} (b) shows the hole-doping dependence of the energies for $U/t=10$.
We find essential similarity to the case $t'/t=0$, indicating the robust stability of the stripe ground state irrespective of the band structure. 
A quantitative difference is, however, that the stripe states as ground states extends in a wider region $0.1<\delta <0.5$. 
Moreover, the ground state at $\delta=0.125$ shows C4S8 order, which is consistent with the experiments of La-based cuprates\cite{Tranquada1995,Tranquada1996}.
The charge and spin configurations of the C4S8 ground state at $\delta=0.125$ are shown in Figs. \ref{config} (c) and (d), respectively.
This C4S8 state stably exists as the ground states for $0.11 \leq \delta \leq 0.15$ although it severely competes with other stripe order such as C3S3 and C5S5. 
The locking of stripe period has been recently observed in the scanning-tunneling-microscope experiment combined with phase resolved electronic structure visualization technique\cite{Mesaros2016a}.
Below $\delta \sim 0.1$, a PS between antiferromagnetic and stripe states occurs as with the case of $t'/t=0$.

\subsection{Spin, charge and superconducting orders}
\begin{figure}[htbp]
  \begin{center}
   \includegraphics[width=82mm]{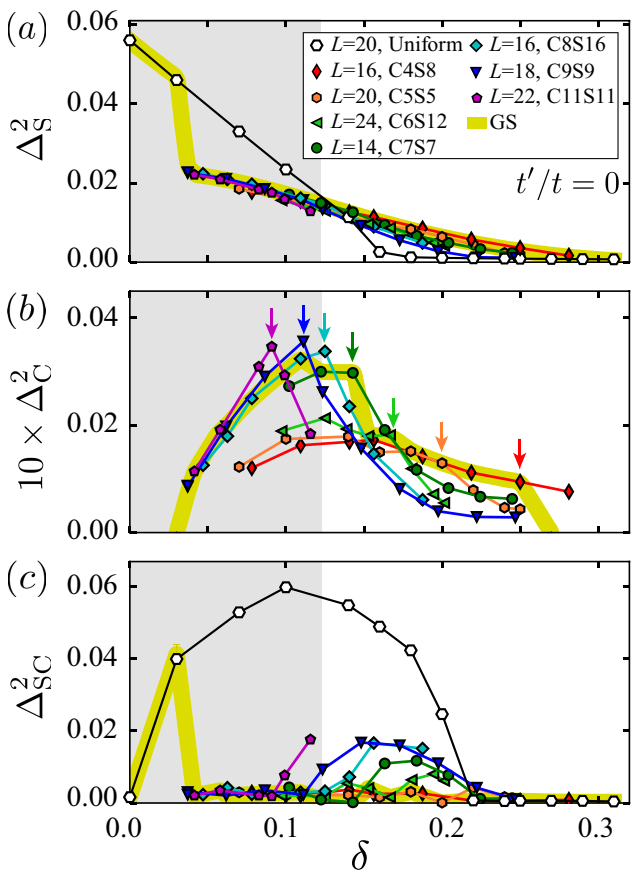}
  \end{center}
  \caption{
    (Color online) $\delta$-dependence of (a) $\Delta_{\rm S}^2$, (b) $\Delta_{\rm C}^2$ and (c) $\Delta_{\rm SC}^2$ for $U/t=10$ and $t'/t=0$.
    Notations are the same as in Fig. \ref{ene_tp0.0} (a).
    Enlarged view for $\Delta_{\rm SC}^2$ will be shown in Appendix \ref{ap:enlarge}.
  }
  \label{phys_tp0.0}
\end{figure}

The $\delta$-dependence of $\Delta_{\rm S}^2$ and $\Delta_{\rm C}^2$ for $t'/t=0$ are shown in Figs. \ref{phys_tp0.0} (a) and (b), respectively.
We see that $\Delta_{\rm S}^2$ decreases as $\delta$ increases.
On the other hand, $\Delta_{\rm C}^2$ has a dome structure around the maximum at $\delta \sim 0.1$.
The dome-like stripe order exists even after the extrapolation to the thermodynamic limit as shown in Appendix \ref{ap:size}.
 
Figure \ref{phys_tp0.0} (c) shows $\delta$-dependence of $\Delta_{\rm SC}^2$.
We see that $\Delta_{\rm SC}^2$ in the stripe states is substantially smaller than those of charge uniform states.
The previous VMC study showed that the strong superconductivity obtained by assuming the charge uniformity emerges in accord with the region of the PS, and therefore is mostly preempted by the PS\cite{Misawa2014a}.
  In the present study, we have shown that if microscopic inhomogeneity is allowed, large portion of the PS is compromised by the formation of stripes. 
The superconductivity is anyhow weakened by the stripe formation, because of its character, where carrier rich strips are weakly coupled by the Josephson tunneling.  
However, it should be remarked that the uniform strongly SC state also survives as an excited state with the excitation energy in the order of $0.01t$ (in the cuprate scale $\sim 10-100$K) as one sees in Figs.~\ref{ene_tp0.0} (a) and (b).
$\Delta_{\rm SC}^2$ in the uniform state has a dome structure~\cite{Misawa2014a} similar to $\Delta_{\rm C}^2$ in the ground state as one sees in Figs.~\ref{phys_tp0.0}(b) and (c). 

\begin{figure}[htbp]
  \begin{center}
   \includegraphics[width=82mm]{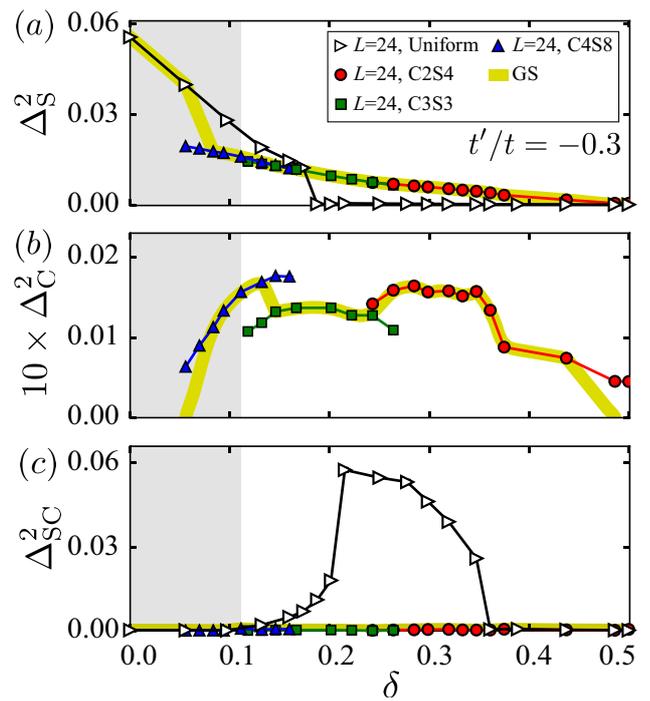}
  \end{center}
  \caption{
    (Color online) $\delta$-dependence of (a) $\Delta_{\rm S}^2$, (b) $\Delta_{\rm C}^2$ and (c) $\Delta_{\rm SC}^2$ for $U/t=10$ and $t'/t=-0.3$.
    Notations are the same as in Fig. \ref{ene_tp0.0} (b).
    Enlarged view for $\Delta_{\rm SC}^2$ will be shown in Appendix \ref{ap:enlarge}.
  }
  \label{phys_tp-0.3}
\end{figure}

Figure \ref{phys_tp-0.3} plots physical quantities for the case of $t'/t=-0.3$, which are again similar to the case of $t'/t=0$.
Note that the stripe order parameters remain finite in the thermodynamic limit below $\delta \sim 0.4$ (see also Appendix \ref{ap:size}).
However, in the experiments, the stripe state has been observed only below $\delta \sim 0.2$\cite{Comin2016}.
This discrepancy will be discussed later.

%
\subsection{Interaction-dependence for $t'/t=-0.3$}
\begin{figure}[htbp]
  \begin{center}
   \includegraphics[width=82mm]{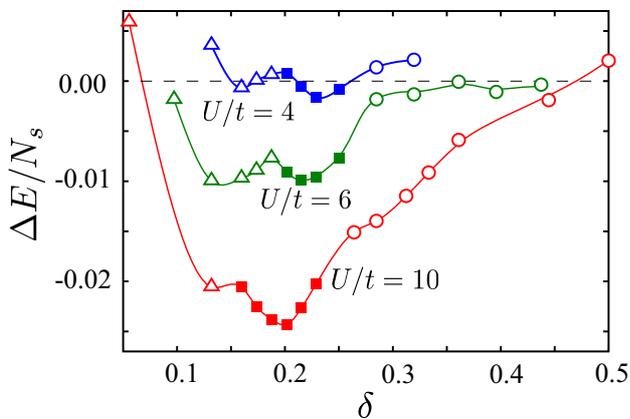}
  \end{center}
  \caption{(Color online) 
    Interaction dependence of the stability of uniform and inhomogeneous states (the energy difference $\Delta E = E_{\rm stripe}-E_{\rm uniform}$) for $t'/t=-0.3$.
    Here, $E_{\rm stripe}$ and $E_{\rm uniform}$ are the energies of stripe and uniform states, respectively.
    Circle, square and triangle symbols show the energies of  C2S4, C3S3, and C4S8 stripes, respectively. 
    Red, green and blue symbols represent $\Delta E$ for $U/t=10$, $6$, and $4$, respectively. 
    Curves are guides for the eyes.
  }
  \label{diff_tp-0.3}
\end{figure}

%
%
Finally, we show the interaction dependence of the energy difference between the uniform and inhomogeneous states for $t'/t=-0.3$ in Fig. \ref{diff_tp-0.3}. 
The stripe states are the ground states above $U/t \sim 4$ and the stripe phase extends with the increase in $U$.
For $U/t = 6$, the stripe and the uniform strongly SC states are nearly degenerate around $\delta \sim 0.3$.
The stripe and uniform SC order parameters become smaller compared with those for $U/t=10$ but the $\delta$-dependence is similar, and we do not find a clear indication of PS. (See Appendix \ref{ap:u6}.)
At $U/t=4$, the charge uniform state is nearly degenerate with the stripe state but the order parameters for the stripe and SC are all nearly zero in the both states, implying that the ground state is a paramagnetic metal. 
Although the stability changes, the stripe and SC orders have similar trend in the dependences on $U$ and $\delta$.

\section{Discussion}\label{discussion}
The same trend between the stripe and SC orders is naturally understood because the emergent and strong effective attractive interaction of carriers, which arises from the originally repulsive interaction, generates both of the order.  
The stripe as a consequence of aggregation of carriers in the real space, and the strong coupling superconductivity both requires strong effective attraction of carriers. 
The effective attraction may have both static and retarded pieces. 
It is possible that the latter may be contributed from bosonic glues including spin fluctuations\cite{Maier2008,Kyung2009,Civelli2009,Scalapino1986,Miyake1986} and reinforced by hidden fermion excitations\cite{Sakai2016a,Sakai2016b}.
The static effective attraction is a direct consequence of the negative quadratic coefficient $b<0$  in the energy expansion $E=E_0+a\delta +b\delta^2+\cdots$ as seen in Figs.\ref{ene_tp0.0} (a) and (b).
$b<0$ is caused by the Mottness, where the kinetic energy decreases nonlinearly upon doping~\cite{Misawa2014a}.

In the presence of realistic values of $t'/t$ and $U/t$ for the cuprates, our calculations show the severe competition among stripe states with $l_c=3-7$ below $\delta \sim 0.2$. 
The charge wavelengths $l_c=3-7$ have been observed in a number of cuprates for $0.05 \lesssim \delta \lesssim 0.2$\cite{Tranquada1995,Tranquada1996,Mesaros2016a,Ghiringhelli2012,Tabis2014,Comin2015,Comin2015a,Forgan2015,Peng2016,Hucker2013,Yamada1998,Campi2015,Fink2011}.
The wavelength of charge $l_c=4$ is consistent with the observations not only in La-based cuprates\cite{Tranquada1995,Tranquada1996} but also in a Bi-based cuprate\cite{Mesaros2016a}. 
The charge inhomogeneity with $l_c=5-7$ has been observed in La-based cuprates below $\delta \sim 0.1$\cite{Yamada1998,Hucker2013,Fink2011}.
The wavelength $l_c=3$ is close to the experimental observations for a Y-based cuprate\cite{Ghiringhelli2012,Tabis2014,Comin2015,Comin2015a,Forgan2015,Peng2016}.
The charge wavelengths observed in a single-layered Hg-based cuprate are $l_c\approx 3.58$\cite{Tabis2014} and $4.35$\cite{Campi2015}, which is located within $l_c=3-5$.
Recent first-principles studies have shown that the single-layered Hg-based cuprate has weaker effective Coulomb interactions than the single-layered La-based cuprate\cite{Jang2015,HirayamaMisawaYamajiImada2017}.
Our results support these studies because the inhomogeneities become weaker with weakening of the interaction, which is consistent with the experiments where the charge order in the Hg-based cuprate is much weaker than that in the La-based cuprate\cite{Hucker2013,Fink2011,Tabis2014}.

The parameter values $t'/t=-0.3$ and $U/t=10$ were proposed as realistic values for the cuprates\cite{Hybertsen1990,Tsutsui1999,HirayamaMisawaYamajiImada2017}. 
However, our results show that the stripe phase is extended in a much wider range of $\delta$ compared with the experiments. 
On the other hand, by weakening $U/t$, the stripe order parameters and the energy difference between the stripe states and the uniform SC state becomes small.
These results imply that an appropriate description of single-band effective hamiltonians for the cuprates is found in the region of intermediate on-site interactions rather than the strong coupling region at least in terms of the stability of the stripe and SC phases. 

The reason why the $d$-wave SC ground state does not clearly appear in contradiction to the experimental results is speculated to be the oversimplification of the Hubbard models we studied.
As recent numerical results are consistent with each other\cite{Zheng2017}, the discrepancy does not seem to originate from the limitation of the accuracy of our calculations (see also the last 
paragraph of this section).
In order to make a more quantitative and reliable comparison with experiments beyond our present analysis, we should analyze the {\it ab initio} effective Hamiltonians, which include long-range Coulomb interactions and hopping integrals and, if necessary, the electron-phonon coupling missing in the simplified Hubbard model.
For example, in the {\it ab initio} single-band effective Hamiltonian for the Hg-based cuprate\cite{HirayamaMisawaYamajiImada2017}, the nearest-neighbor Coulomb interaction is about 20\% of the on-site interaction. 
The third-nearest-neighbor hopping $t''$ in the effective Hamiltonian has also a non-negligible value of $t''/t \sim 0.15$\cite{HirayamaMisawaYamajiImada2017}.
A tiny energy difference between the superconducting and stripe states is subject to be easily reversed by such realistic factors.
We are now at the stage that allows quantitative comparisons between model calculations and the experimental results, because of the achieved accuracy of the solver.
The origin of the quantitative discrepancy will be discussed elsewhere based on first-principles studies.

One may be concerned about the accuracy of the present calculation.
However, our trial wave function can be systematically improved by using methods such as the power Lanczos and/or tensor network\cite{Heeb1993,Neuscamman2012,Chou2012,Sikora2015,Zhao2017}.
These additional refinements indeed lower the energies. However, the energies are nearly equally lowered among competing states, and other physical quantities such as stripe and superconducting orders only slightly change\cite{Misawa2014a,Zhao2017}. 
(See also Appendix \ref{ap:lan}.)

\section{Summary}
Our VMC calculations show stripe ground states of the Hubbard models irrespective of the amplitude of the next nearest neighbor hopping. 
Its stability and stripe order parameter substantially increases with increasing $U$ in the strong coupling region beyond $U/t=5$ and becomes extended in a wider range of hole doping concentration with a dome-like $\delta$ dependence. 
With increasing hole doping, the stripe period decreases.
The stripe period is roughly proportional to the mean hole distance for $t'/t=0.0$, whereas it is not for $t'/t=-0.3$.
This detailed difference may be ascribed to the difference in the Fermi surface nesting vectors especially in the antinodal region. 
This issue will be studied in future studies.
The period at $t'/t=-0.3$ agrees with that observed in the experiments at $\delta=0.125$.

In the static stripe ground states, the superconductivity is substantially suppressed. 
On the other hand, metastable excited states with the uniform and strongly SC order, whose excitation energy is tiny ($\sim 0.01t$), appear with dome-like $\delta$ dependence similarly to the dome of charge stripe order. 
The superconducting order, in both excited and ground states decreases for smaller $U/t$ and numerically invisible for $U/t \lesssim 4$ which again has trend essentially similar to the charge order.

The same trend between the SC and stripe states and their severe competition are a consequence of the strong effective attraction originating from the strong repulsive interaction.  
  Understanding their common route and distinctions revealed here will help designing ways of suppressing the stripe and stabilizing the SC state simultaneously.  
  Some attempts were already made\cite{Misawa2016, Ido2017}, and extensive studies along this line are intriguing challenging issues in the future.

An interesting future issue is to more quantitatively analyze effective low-energy hamiltonians of the cuprates obtained from {\it ab initio} calculations~\cite{HirayamaMisawaYamajiImada2017} to understand the mechanisms and materials dependence in the light of the present severe competitions. 
In particular, the validity of the single-band description has to be seriously examined because the present elucidation suggests a weaker correlation than the parameters proposed in the literature~\cite{HirayamaMisawaYamajiImada2017} if one sticks to the single-band description.

%
\begin{acknowledgments}
%
The authors thank the Supercomputer Center, the Institute for Solid State Physics, the University of Tokyo for the facilities. 
The calculations were performed by using the open-source software mVMC\cite{mVMC2017}.
We thank the computational resources of the K computer provided by the RIKEN Advanced Institute for Computational Science through the HPCI System Research project, as well as HPCI Strategic Programs for Innovative Research (SPIRE), the Computational Materials Science Initiative (CMSI), and Social and scientific priority issue (Creation of new functional devices and high-performance materials to support next-generation industries; CDMSI) to be tackled by using post-K computer, under the project number hp130007, hp140215, hp150211, hp160201, and hp170263 supported by Ministry of Education, Culture, Sports, Science and Technology, Japan (MEXT) .
This work was also supported by Grant-in-Aids for Scientific Research (No. 22104010, No. 22340090 and No. 16H06345 ) from MEXT.  
KI was financially supported by Grant-in-Aid for JSPS Fellows (No. 17J07021) and Japan Society for the Promotion of Science through Program for Leading Graduate Schools (MERIT).
%
\end{acknowledgments}

\bibliographystyle{prsty}
\bibliography{reference}
%
%
\clearpage
\appendix
\section{Enlarged view of $\delta$ dependence of $\Delta_{\rm SC}$}\label{ap:enlarge}
\begin{figure}[htbp]
  \begin{center}
   \includegraphics[width=82mm]{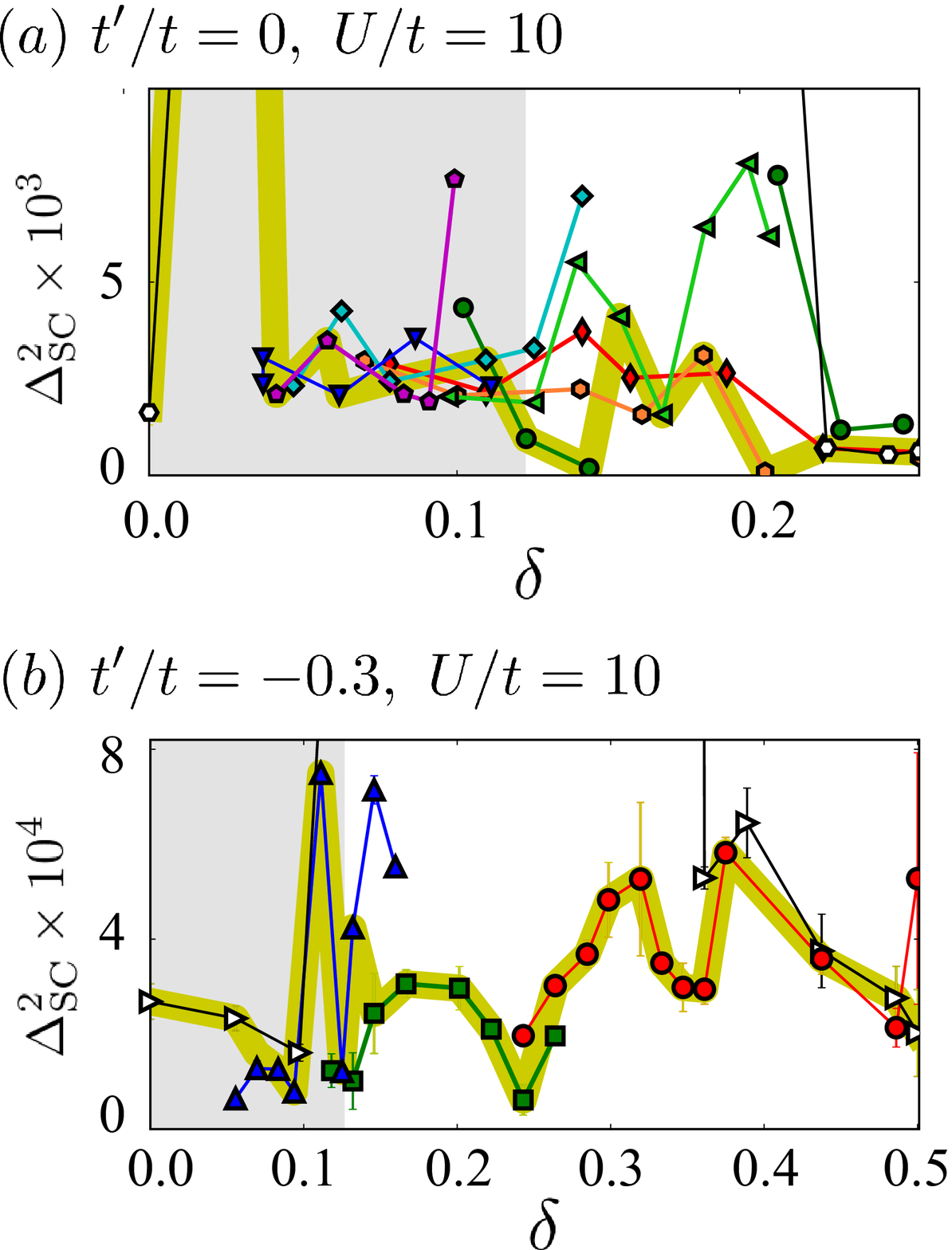}
  \end{center}
  \caption{(Color online) Doping concentration dependence of superconducting order parameter for $U/t=10$ at (a) $t'/t=0$ and (b) $t'/t=-0.3$. 
    Notations in the panels (a) and (b) are the same as Figs. 4 and 5, respectively.
  }
  \label{enlarge}
\end{figure}

Figures \ref{enlarge} (a) and (b) show the enlarged views of Figs. 2 (c) and 2 (c) in the main text which plot the hole concentration dependence of the superconducting order parameters for $t'/t=0$ and $t'/t=-0.3$, respectively. 
The maximum value of the SC order paramters for the ground states is the order of $10^{-4}-10^{-3}$ at most, which is much smaller than that of the uniform excited state.

%
\section{Size-dependence of stripe and superconducting order parameters for stripe states}\label{ap:size}
\begin{figure}[htbp]
  \begin{center}
   \includegraphics[width=80mm]{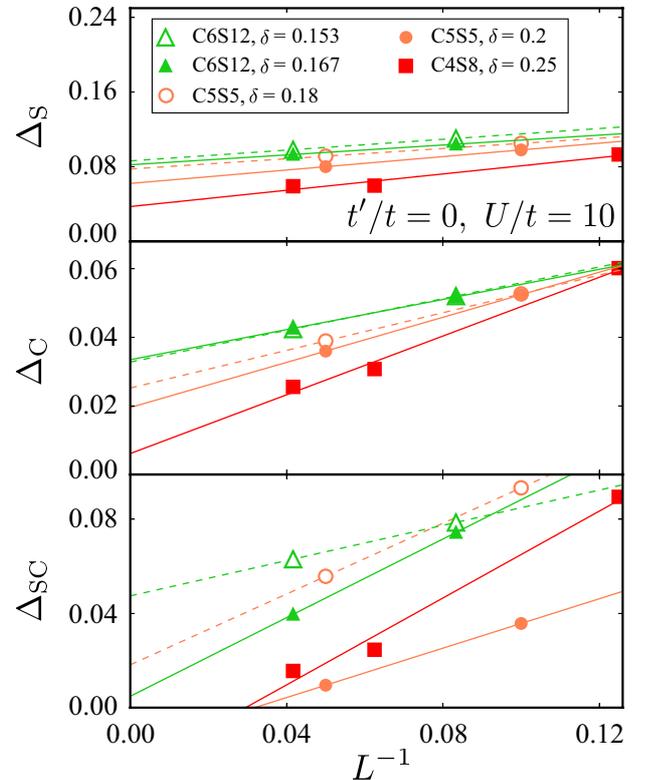}
  \end{center}
  \caption{(Color online)  System-size dependece of order parameters for $t'/t=0$ and $U/t=10$. 
  In the legend, types of quantum states and hole concentrations are described.
  Solid symbols correspond to the commensurate fillings in which 
one hole fills per one charge-stripe unit cell.
  Solid and dashed lines represent the linear-extrapolation fittings by $a+bL^{-1}$.
  }
  \label{size_tp0.0}
\end{figure}

%
To clarify the thermodynamic properties of the ground states, we show the size-dependence of physical quantities for $t'/t=0$ and $U/t=10$ within the stripe ground state at several doping concentrations in Fig. \ref{size_tp0.0}. 
Here, following the convention in the literature~\cite{Assaad2013}, we estimated the extrapolated order parameter $\Delta$ by fitting the several points with $a+bL^{-1}$. Even when we employ the scaling $a'+b'L^{-1/2}$, the results do not essentially change. 
Figure \ref{size_tp0.0} shows that both the spin and charge order parameters remain finite even after the extrapolations below $\delta \sim 0.2$.
At commensurate fillings, one hole fills in a one charge wavelength, i.e. $\delta = 1/l_c$. 
The bottom panel of Fig.\ref{size_tp0.0} shows, in the thermodynamic limit, clear stronger suppression of long-range superconducting order at commensurate fillings $\delta = 1/l_c$ than the case $\delta \neq 1/l_c$ incommensurate to the stripe period.
In the latter incommensurate fillings, the superconducting order likely remains nonzero in the thermodynamic limit. 

\begin{figure}[htbp]
  \begin{center}
   \includegraphics[width=80mm]{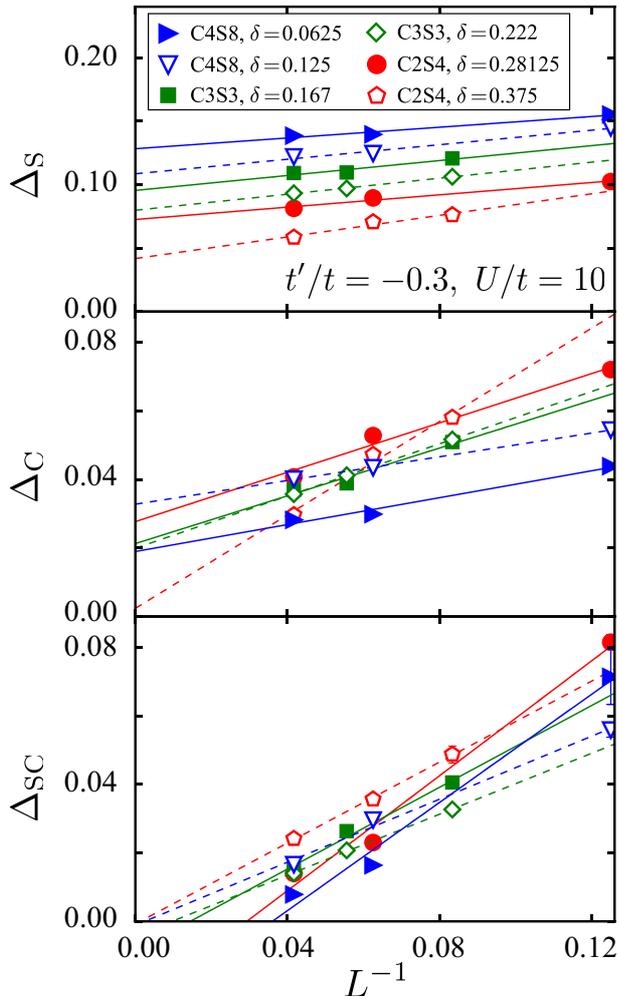}
  \end{center}
  \caption{(Color online)  System-size dependece of order parameters for $t'/t=-0.3$ and $U/t=10$. 
  In the legend, types of quantum states and hole concentrations are described.
  Solid and dashed lines represent the linear-extrapolation fittings.
  }
  \label{size_tp-0.3}
\end{figure}

We also show size-dependences of physical quantities for $t'/t=-0.3$ in Fig.~\ref{size_tp-0.3}.
As we mentioned in the main text, the extraporated values of stripe orders have nonzero values below $\delta \sim 0.4$.
On the other hand, we do not find any non-positive extrapolated values of the SC order parameter at this stage, which is different from the case of $t'/t=0$.
  To understand this difference, we need further analysis of the size dependence of the SC order parameter and its doping dependence in the thermodynamic limit for both $t'/t=0$ and $t'/t=-0.3$, but it is left for a future study.

%
\section{Physical quantities for $t'/t=-0.3$ and $U/t=6$}\label{ap:u6}
\begin{figure}[htbp]
  \begin{center}
   \includegraphics[width=82mm]{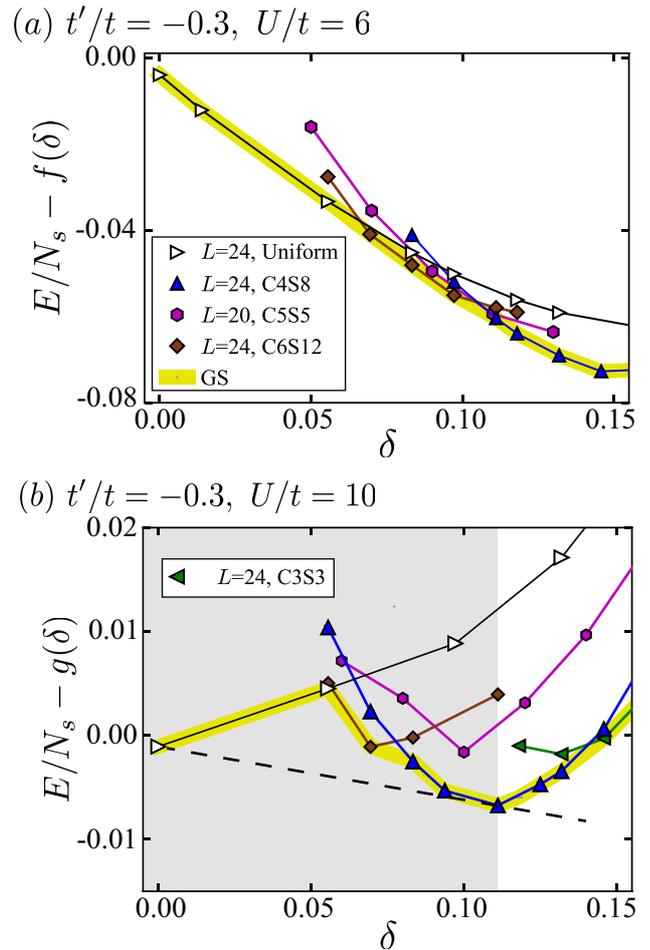}
  \end{center}
  \caption{(Color online) Doping dependence of the energy of several different states for $t'/t=-0.3$ below $\delta=0.15$. 
  We set $f(\delta)=-0.8\delta-0.640$ and $g(\delta)=-1.7\delta-0.4211$.
  Types of states and system sizes are described in the legend.
  For clarity we draw yellow thick line for the energies of the ground states for $L=24$. 
  Dashed black line and gray region show the tangent line of the energy curve and PS, respectively.
  }
  \label{ene_u6tp-0.3}
\end{figure}

%
%
Figure \ref{ene_u6tp-0.3} compares the hole-doping dependence of the energies between $U/t=6$ and $U/t=10$ below $\delta \sim 0.15$.
We do not find an evidence for the PS between the antiferromagnetic state and the stripe state at $U/t=6$,
where a tangent line from $\delta=0$ to the energy curve cannot be drawn, distinctly from the case $U/t=10$.

\begin{figure}[htbp]
  \begin{center}
   \includegraphics[width=84mm]{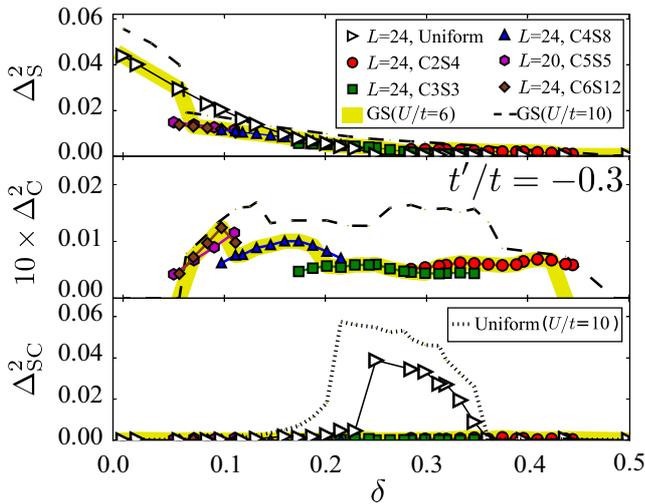}
  \end{center}
  \caption{(Color online) Doping dependence of order parameters for (a) spin and (b) charge stripes for $U/t=6$ and $t'/t=-0.3$. 
    Dashed and dotted curves represent the results of the ground states and the charge uniform state for $U/t=10$ for comparison, respectively.
    Notations are the same as in Fig. 5.
  }
  \label{phys_tp03_u6}
\end{figure}

\begin{figure}[htbp]
  \begin{center}
   \includegraphics[width=84mm]{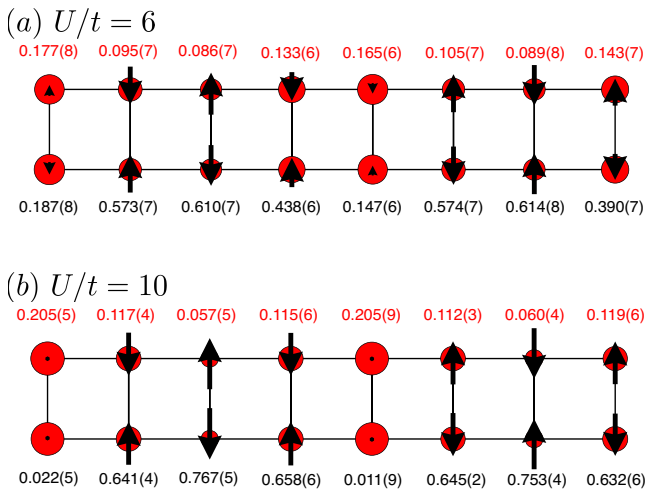}
  \end{center}
  \caption{(Color online) Spin density along $z$-direction $S^z_i=\braket{n_{i\uparrow}-n_{i\downarrow}}$ and hole density $1-\braket{n_i}=1-\braket{n_{i\uparrow}+n_{i\downarrow}}$ for $t'/t=-0.3$ at $\delta=0.125$ for the ground state with C4S8 for $L=24$.
  The radius of every red circle is propotional to the hole density $1-n_i$. 
  The length of every black arrow is proportional to the amplitude of the spin density $|S^z_i|$. 
  The values of $|S^z_i|$ and $1-n_i$ averaged over $y$-direction are shown above and below the plots, respectively.
  Note that the simulations were performed for finite size systems. Nevertheless, the variational wavefunctions show translational symmetry breaking when the momentum projection is not operated. Although a better ground-state wavefunction is obtained after the momentum projection,  the overlap of the two functions spatially translated each other is negligible in the size $L=24$ and the orderparameter is expected to be close to the thermodynamic limit.
  }
  \label{local_tp03}
\end{figure}

Figure \ref{phys_tp03_u6} plots the $\delta$-dependence of the spin, charge and superconducting order parameters for $U/t=6$ and $t'/t=-0.3$.
  We see that the results are qualitatively similar to the case of $U/t=10$, but the stripe order parameters become smaller.
This means that the inhomogeneity is weakened by the decrease of the on-site interaction.
This tendency is also seen in Fig. \ref{local_tp03}, where the electron distribution in real space is depicted. 
The superconductivity in the uniform excited states has the same trend as the case of the stripe orders. 
At $U/t=4$, the stripe and superconducting orders are scaled to zero within the numerical accuracy.

%
\section{Power lanczos method}\label{ap:lan}
%
The power lanczos method is one of the systematic ways to improve a trial wave function in the VMC method\cite{Heeb1993}.
In the $N$-th power Lanczos method, we multiply the Hamiltonian to the optimized trial wave function $\ket{\psi_{\rm opt}}$ as
\begin{eqnarray}
  \ket{\psi^{(N)}} =&& \left(1+\sum_{n=1}^{N}\alpha_n \mathcal{H}^n \right) \ket{\psi_{\rm opt}},
\end{eqnarray}
where $\alpha_n$ are the variational parameters.
We use the 1st step Lanczos method ($N=1$) since the numerical costs grow exponetially with increasing $N$.

\begin{figure}[htbp]
  \begin{center}
   \includegraphics[width=82mm]{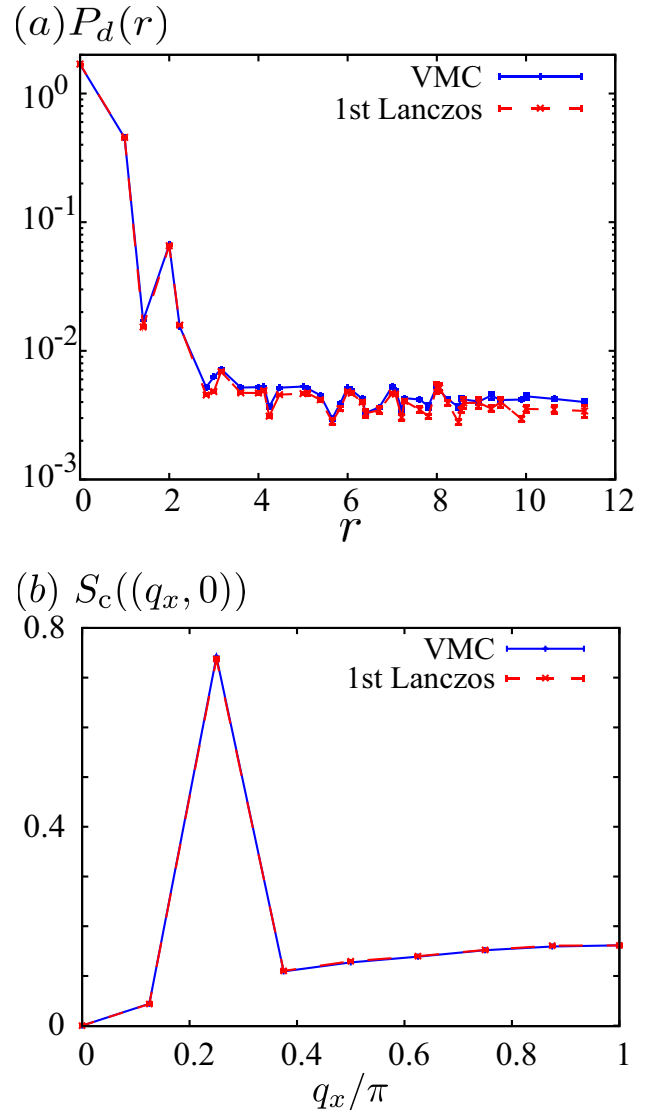}
  \end{center}
  \caption{(Color online) Superconducting correlation function $P_d(r)$ (a) and charge structure factor $S_{\rm c}(\bm{q}_{\rm peak})$ at $\bm{q}_{\rm peak}=(q_x,0)$ (b) of the C8S16 state for $L=16$, $U/t=10$ and $t'/t=0$ at $\delta \approx 0.11$ . 
    Blue line and red dashed line are the results obtained by using the VMC method and the 1st step Lanczos method, respectively.
  }
  \label{phys_lanczos}
\end{figure} 

Table \ref{ene_ls} shows the energies of competing states for various doping concentrations $\delta$.
The Lanczos method improves the energies of competing states but does not change character of the ground states and only slightly alters physical properties as below.
We have checked the effects by the Lanczos operation to other physical quantities such as the superconducting correlation function and the charge structure factor. 
However, these are only slightly changed as shown in Fig. \ref{phys_lanczos}.
 
\begin{table}[htbp]
  \caption{ Energies per site of competing states obtained from the VMC and 1st Lanczos calculations for several system sizes $L$ and the hole-doping concentrations $\delta$ at $U/t=10$ and $t'/t=0$.
    The number in brackets represents the error on the last digit.
  }
    \begin{tabular}{ccc|cc} \hline  \hline
      $L$ & $\delta$ & state & VMC & 1st Lanczos  \\ \hline 
      20 & 0.180 & Uniform & -0.7384(2) & -0.7591(4)\\ 
      20 & 0.180 & C5S5 & -0.74820(4) & -0.7639(8)\\ \hline
      14 & 0.143 & Uniform & -0.6665(5)& -0.6900(7)\\ 
      14 & 0.143 & C7S7 & -0.68315(5) & -0.6992(3)\\ \hline
      16 & 0.109 & Uniform & -0.60744(9) & -0.6272(4)\\ 
      16 & 0.109 & C8S16 & -0.62232(4) & -0.6377(1)\\ \hline  \hline
    \end{tabular}
  \label{ene_ls}
\end{table}

\end{document}